\documentclass[a4paper,11pt]{article}
\pdfoutput=1 

\usepackage{jinstpub} 
\usepackage{subcaption}
\usepackage{url}

\title{\boldmath CALICE SiW ECAL - Development and performance of a highly compact digital readout system}

\author[a]{D. Breton,}
\author[a]{A. Irles,}
\author[a]{J. Jeglot,}
\author[a]{J. Maalmi,}
\author[a,1]{R. P\"oschl \note{Corresponding author.}\note{Talk presented at the International Workshop on Future Linear Colliders (LCWS2019), Sendai, Japan, 28 October-1 November, 2019, C19-10-28, and the International Conference on Calorimetry at the High Energy Frontier (CHEF 2019)  Fukuoka, Japan, 25 November-29 November, 2019, C19-11-25.3.},}
\author[a]{D. Zerwas}

\affiliation[a]{Universit\'e Paris-Saclay, CNRS/IN2P3, IJCLab, 91405 Orsay, France}

\emailAdd{poeschl@lal.in2p3.fr}

\abstract{A highly granular silicon-tungsten electromagnetic calorimeter (SiW-ECAL) is the reference design of the ECAL of the International Large Detector (ILD) concept, one of the two detector concepts for the detector(s) at the future International Linear Collider. Prototypes for this type of detector are developed within the CALICE Collaboration. During the last year a highly compact digital readout system has been built. The system has been used for the first time in a beam test in Summer 2019 at DESY. This article summarises the main features of the system and report on its performance during the beam test.}


\collaboration[c]{on behalf of the SiW ECAL groups within the CALICE Collaboration}

\proceeding{LCWS19 and CHEF2019\\
  when 28/10/19 - 2/11/19 and 25/11/19 - 29/11/19\\
  where Sendai/Japan and Fukuoka/Japan}

\begin{document}
\maketitle
\flushbottom

\section{Introduction}
\label{sec:intro}

\begin{figure}[!t]
  \centering
 \includegraphics[width=0.4\textwidth]{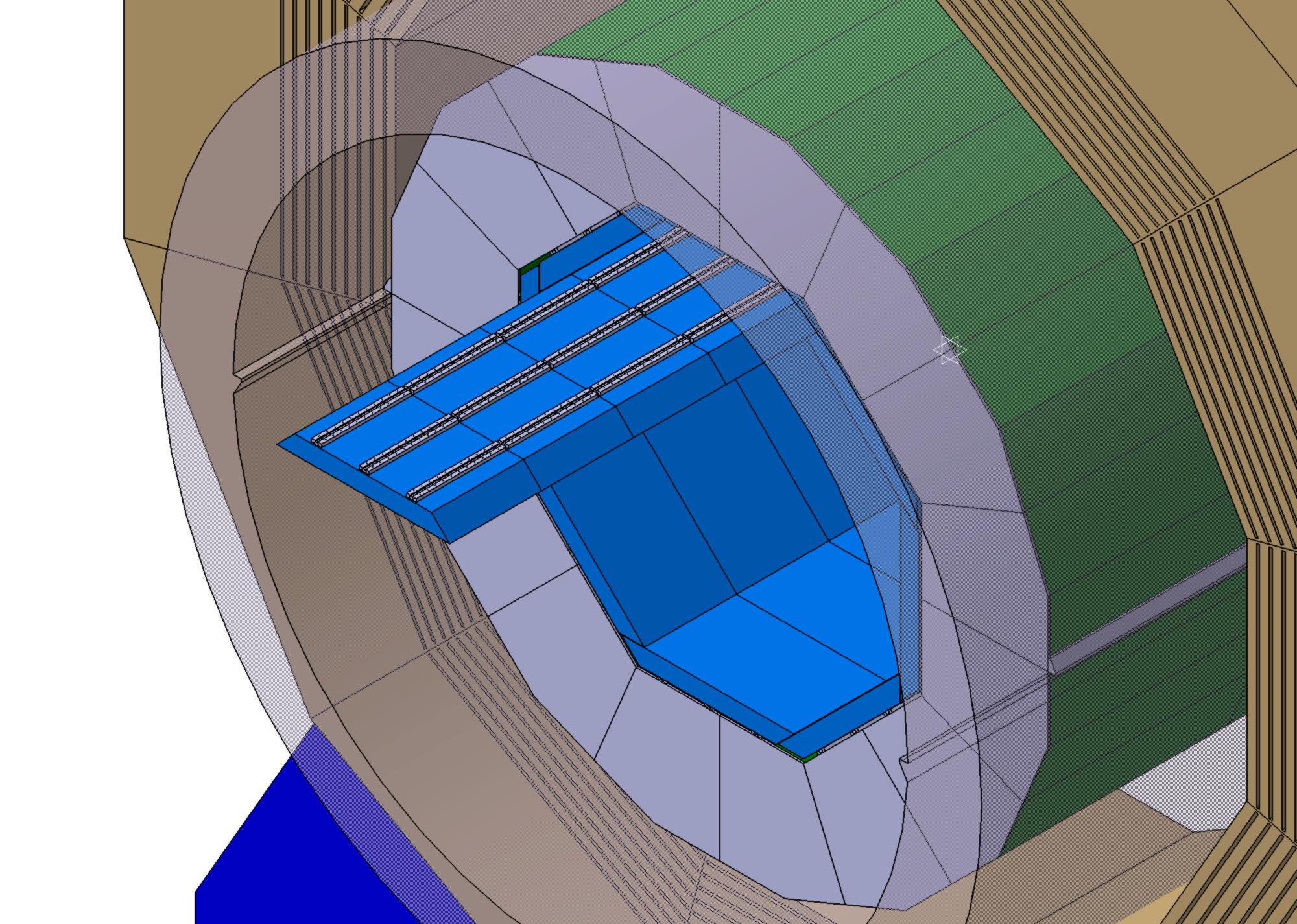}    \includegraphics[width=0.45\textwidth]{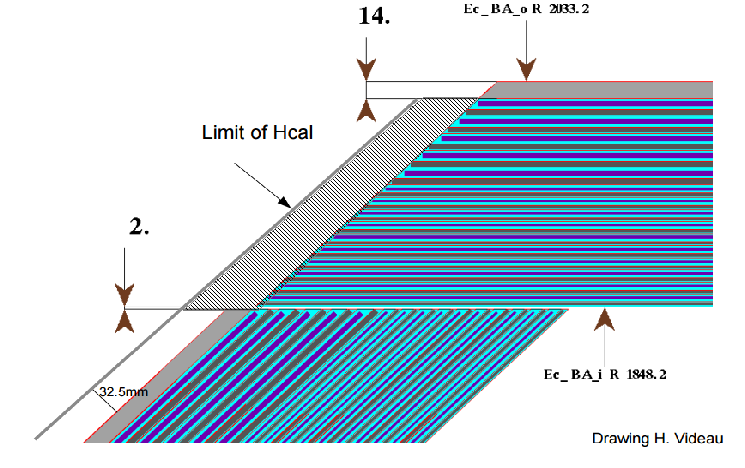} \\ 
\vspace{0.5cm}
 \includegraphics[width=0.85\textwidth]{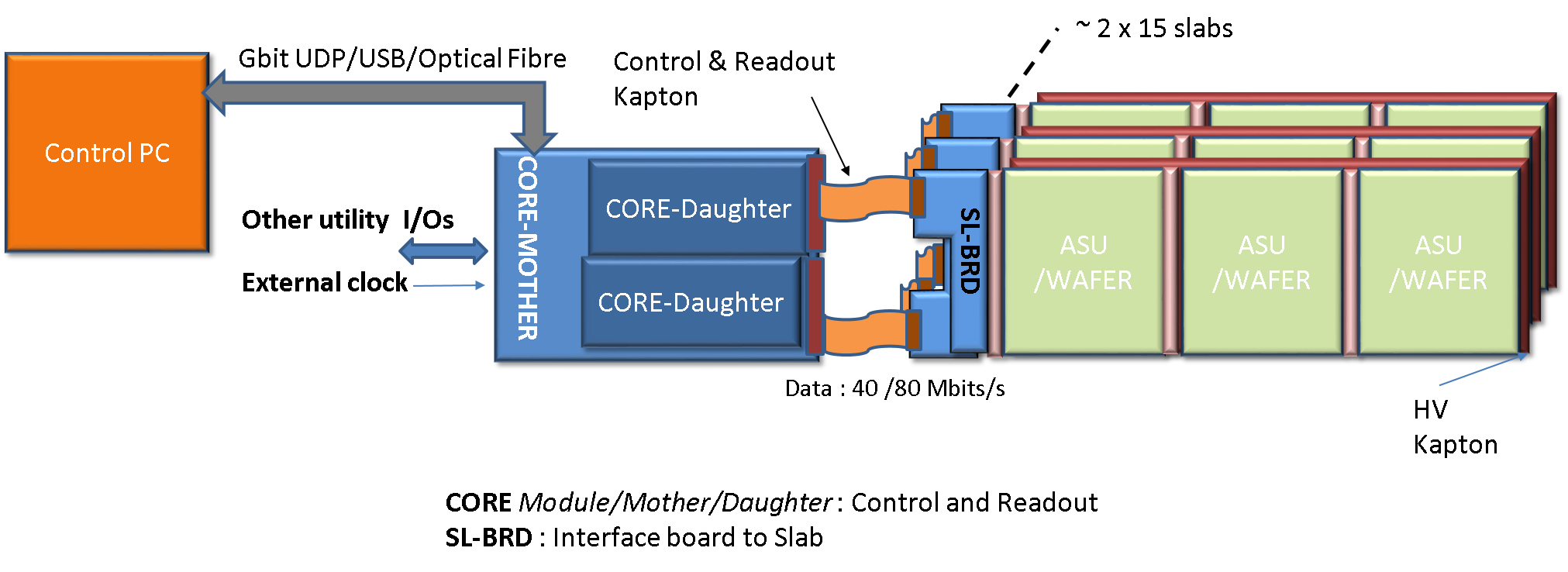}
\caption{\sl \underline{Top Left:} Schematic view on the calorimeters in the ILD Detector. The SIW ECAL in light blue is surrounded by a hadronic calorimeter.
\underline{Top Right:} Schematic view onto a region of the ILD SiW ECAL barrel calorimeter. The hatched area has a width of about 67\,mm that is available for the interface cards of the digital readout of the SiW ECAL layers (indicated in blue).  
\underline{Bottom:} Architecture of the readout system. The SiW ECAL layers are organised in terms of up to ten Active Signal Units (ASU). The readout system starts at the left of the actual layers.  
}
\label{fig:ild-ecal}
\end{figure}

High precision measurements in modern particle physics detectors require a close-to full coverage of the solid angle to ensure a large lever arm for angular distributions or to enhance sensitivity for processes with missing energy. This design requirement implies space- and power-economic solutions for services such as readout or detector control. This article presents a newly developed digital readout system for a technological highly-granular silicon-tungsten electromagnetic calorimeter, SiW ECAL, constructed and operated by the CALICE Collaboration . The dimensions and design specifications are oriented at the needs of the SiW ECAL of the ILD detector~\cite{Behnke:2019sdv}, which is a proposal for a detector at the International Linear Collider ILC~\cite{Behnke:2013xla}.     
The barrel of the SiW ECAL of ILD, see Figure~\ref{fig:ild-ecal}, is organised in trapezoidal modules forming an octant with an inner diameter of around 1850\,mm. The SiW ECAL is surrounded by the hadron calorimeter. As sketched in the top part of Figure~\ref{fig:ild-ecal} around 67\,mm are available for the interface cards to the detector layers and the circuitry for detector control, readout and power supply. The system described in this article replaces the system described in Reference~\cite{Gastaldi:2014vaa} that has been formerly used in a number of beam test campaigns of the SiW ECAL~\cite{Amjad:2014tha,Kawagoe:2019dzh}.

\section{Digital readout system}

The individual elements of the digital readout system sketched in the bottom part of Figure~\ref{fig:ild-ecal}  will be introduced in the following.

\subsection{The SL-Board and the CORE Kapton}
\label{sec:slboard}

\begin{figure}[!t]
  \centering
 \includegraphics[width=0.5\textwidth]{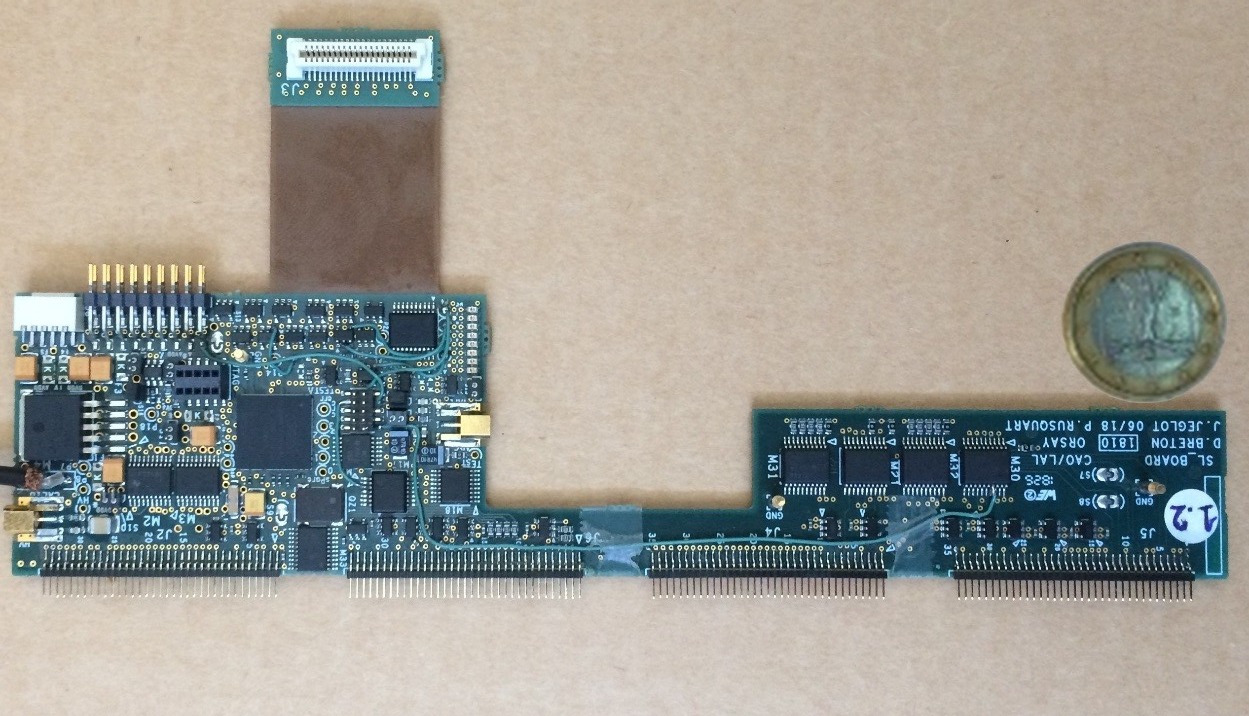}
 \includegraphics[width=0.8\textwidth]{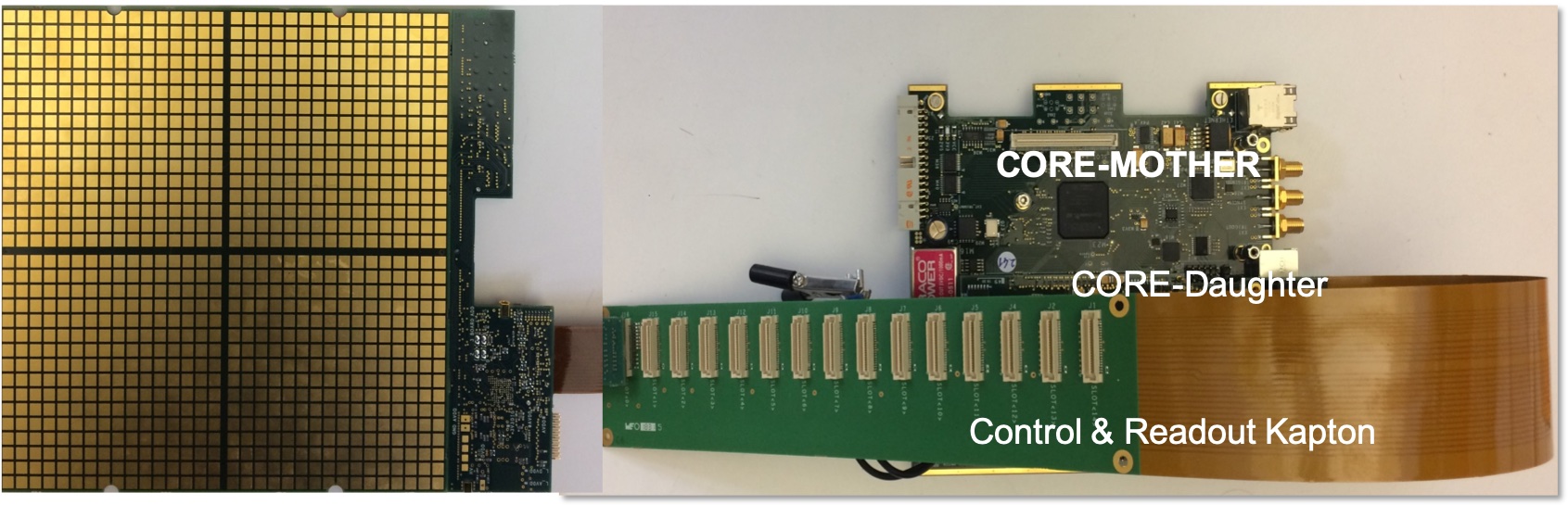}
\caption{\sl \underline{Top:} SL-Board for the digital readout and the power supply of the SiW ECAL layers. \underline{Bottom:} Ensemble of ASU, control and readout kapton and control unit (CORE Mother + CORE Daughter). For details see text.}
\label{fig:SiWECAL_slcards}
\end{figure}

The {\sl SL-Board}, see left part of Figure~\ref{fig:SiWECAL_slcards}, is the sole interface for the $\sim$10000 channels of a detector layer. In case of the SiW ECAL the channels are read by the SKIROC2 and SKIROC2a ASICs of which a detailed description can be found elsewhere~\cite{Callier:2011zz, Kawagoe:2019dzh}. 
The SL-Board is designed to fit the space and mechanical constraints of the SiW ECAL of ILD. Being an effective prolongation of a layer (see Figure~\ref{fig:ild-ecal} bottom for illustration) the SL-Board has a length between 10 and 42\,mm and a width of 180\,mm. The design allows for the integration of a cooling system for the SiW ECAL~\cite{Grondin:2017qzp}. The height of the SL-Board must be contained between 6-12\,mm in order to accommodate 15 SL-Boards above each other in an ILD SiW ECAL barrel module.  
The SL-Board delivers the regulated power including the high voltage for the sensor bias (150-200 V depending on the sensor type and thickness), controls the front-end electronics ASICs and performs the full data readout. 
The SL-Board is based on a MAX10 from ALTERA~\cite{bib:max10}, which is a mix of CPLD and FPGA and features an ADC 
to monitor the power pulsed operation of the detector layers, an option that is available at linear colliders.
It connects to the ASU chain via four surface mounted connectors of 1.5\,mm height and 1\,mm pitch. 
The SL-Board houses a connector towards a FTDI USB module that can be used for standalone and debugging tests. The SL-Board generates all required supplies via regulators from a single 3.4V power input. The current flowing through the board is shared between its own needs and the power delivered to a layer. The consumption of the SL-Board itself is only 200\,mA which corresponds to less than 700\,mW at the 3.4\,V operational voltage. The SL-Board is able to deliver up to 2\,A to a detector layer (mainly on analogue 3.3V), but also to actively limit the layer consumption to 100\,mA in power pulsing mode.


Within a detector system the SL-Board is connected to the outside world through a system dubbed {\sl CORE Kapton} (see Figure~\ref{fig:SiWECAL_slcards}) via an internal kapton layer and a 40-pin connector. The distinguished feature of this solution is that the kapton cable can be guided along the sides of an SiW ECAL module avoiding hence bulky cabling especially since the angle of the barrel module side is 45$^o$, see Figure~\ref{fig:ild-ecal}. The short kapton cables that leave from the SL-Boards provide the necessary flexibility for a mechanically safe and handy connection of the 15 layers that share one CORE Kapton.
The link provided by the CORE Kapton is designed to drive, read out and synchronise up to 15 detector layers. It transmits all the clock
and fast signals and houses the control and readout links. The CORE Kapton interfaces make use of asynchronous serial transmissions, which ease the synchronisation of the detector layers. The speed of the slow control and the individual readout links is 40\,Mbits/s to be compared with the readout of the ASUs constituted by two parallel chains of 5\,Mbit/s.

\subsection{The CORE Module}
\label{sec:core}

\begin{figure}[!t]
  \centering
 \includegraphics[width=0.3\textwidth]{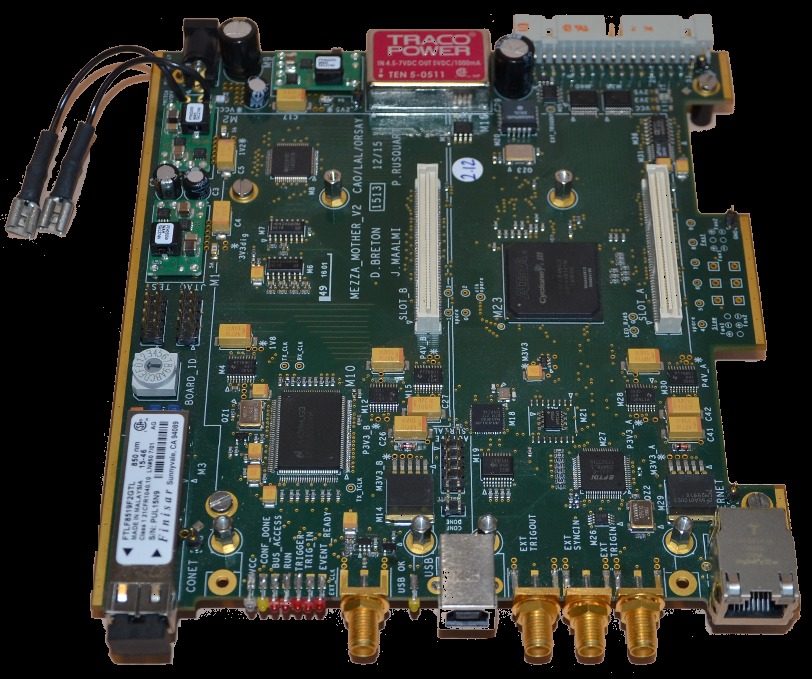}
 \includegraphics[width=0.6\textwidth]{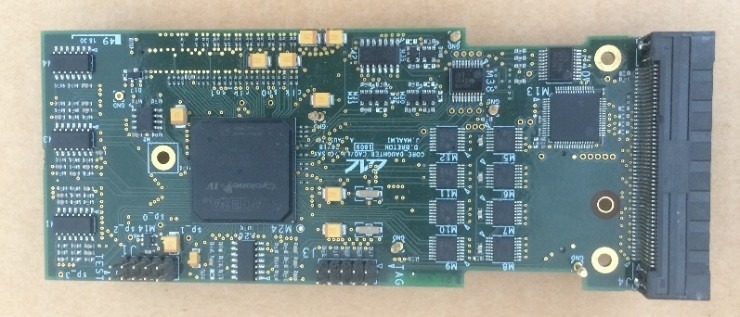}\\
\vspace{0.5cm}
\includegraphics[width=0.45\textwidth]{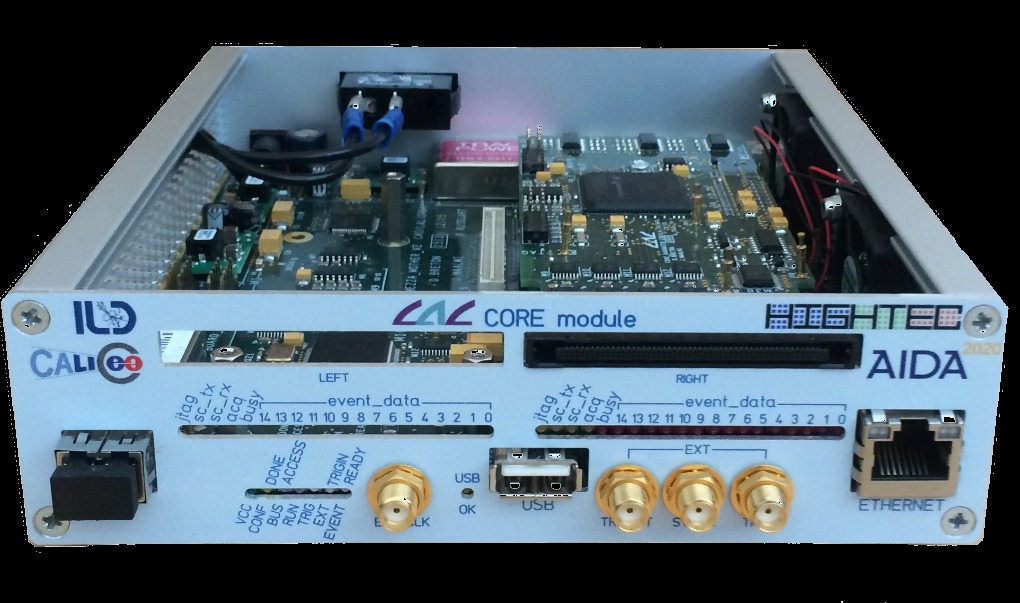}
\caption{\sl \underline{Top:} CORE Mother (left) and CORE Daughter (right) boards that are connected to the SL-Boards. \underline{Bottom:} CORE Module that houses one CORE Mother and two CORE Daughter boards.}
\label{fig:SiWECAL_core}
\end{figure}

The CORE Module consists of the CORE Mother and two CORE Daughter boards. 
Both types are shown in Figure~\ref{fig:SiWECAL_core}. The CORE Mother has been extensively used in worldwide used instruments 
like the WaveCatcher and SAMPIC fast waveform digitisers~\cite{bib:sam-wc}, which have been both developed at IJCLab (formerly LAL Orsay). This control and readout motherboard has been developed for housing up to two mezzanines and it permits separating the acquisition part from the specific front-end part. 
It manages external input and output signals for interfacing or synchronising with other modules. 
The CORE Mother sends common clocks and fast signals
to the CORE Daughters to keep the system synchronised. The control and readout is possible
through USB (2.0) or the custom secured G-bit UDP protocol via a copper or an optical link.
The CORE Daughter has been specifically developed for the SiW-ECAL prototype. It is based on a Cyclone IV FPGA. 
It is the interface between the CORE Mother and the CORE Kapton interface. It houses the second level of event buffers (derandomisers).
The control and readout link between CORE Daughter and CORE Mother is 60\,MBytes/s in case of an USB connection, and 125\,Mbytes/s if the UDP protocol is used. The entire CORE Modules consumes 5\,W.

\subsection{Control and readout Software}
\label{sec:software}

\begin{figure}[!t]
  \centering
\includegraphics[width=0.9\textwidth]{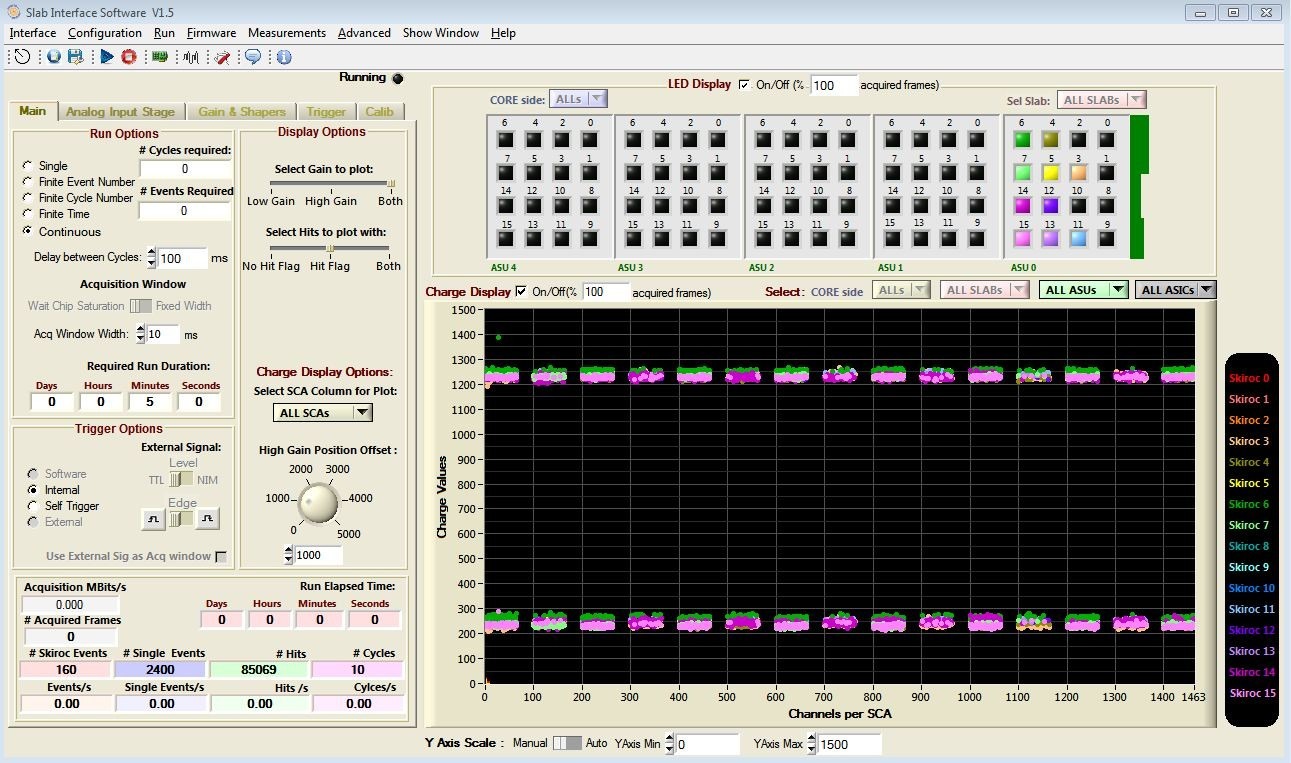} 
\caption{\sl Screenshot of the main window of the software interface to control and readout the modules.}
\label{soft}
\end{figure}

The acquisition software is written in C-Language and developed under LabWindows CVI. It pilotes
the communication through the CORE Kapton or through the FTDI Module directly to the SL-Board. It handles
the control and readout of a whole detector module consisting of two CORE Kaptons connected to 15 SL-Boards 
each and up to five ASUs connected in series to each SL-Board.
The c-functions that handle the communication (readout and configuration) can be used as a library with 
any other program that handles C-Language such as be EUDAQ2~\cite{Liu:2019wim} or Pyrame~\cite{Magniette:2018wdz}.
The software also allows for advanced online commissioning measurements
as such as threshold scans or masking of channels. A screenshot of the main window of the
graphical interface is shown in Figure~\ref{soft}.

\subsection{Beam test performance and outlook}

\begin{figure}[!t]
  \centering
\includegraphics[width=0.8\textwidth]{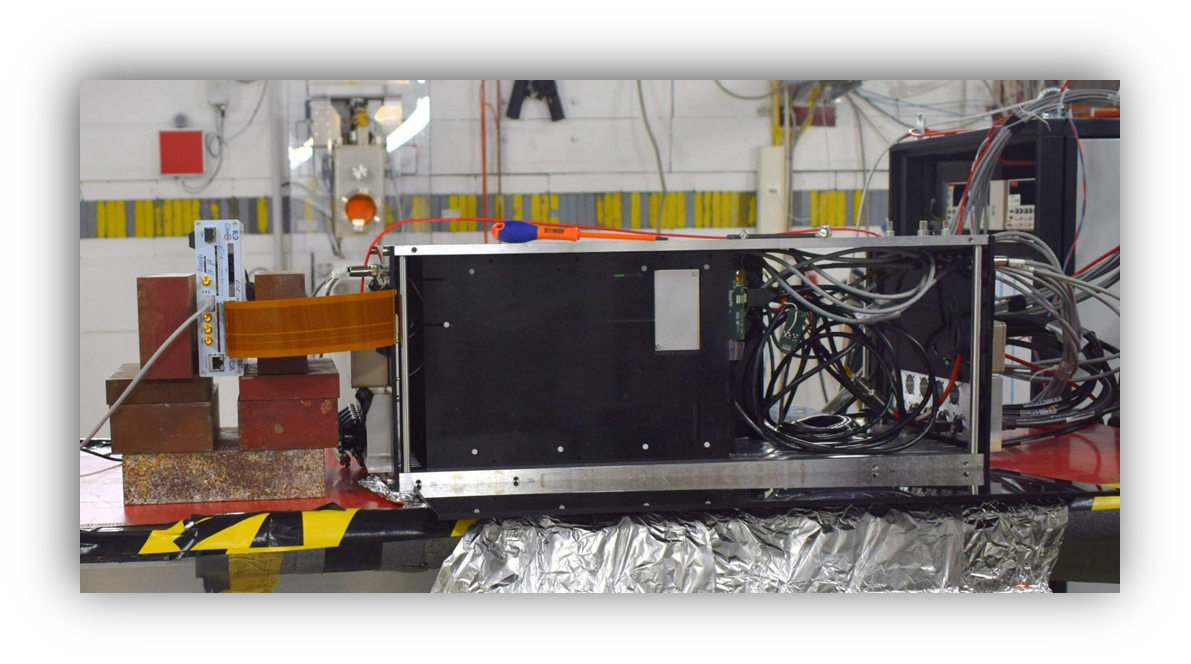} 
\includegraphics[width=0.3\textwidth,angle=90]{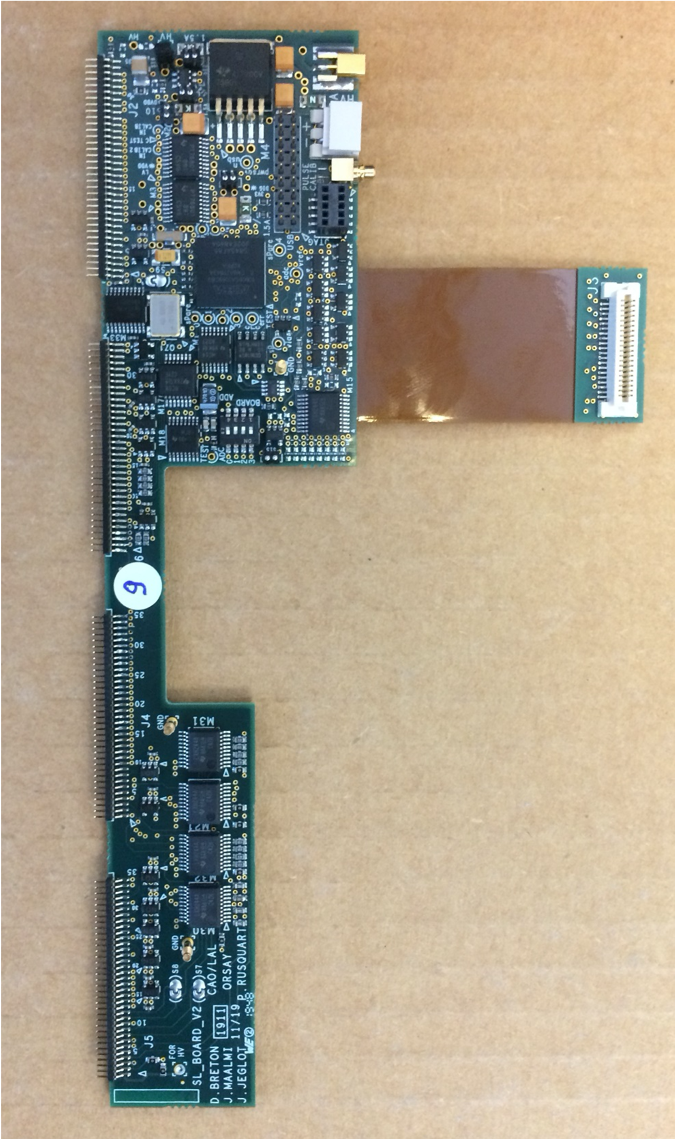} 
\caption{\sl \underline{Top:}Beam test setup at DESY. The flat cable of the CORE Kapton interface leaves the detector on the left hand site and is plugged into the CORE Module. The mechanical housing has been shared with layers using the readout system described in Reference~\cite{Gastaldi:2014vaa} of which the services leave to the right hand side. \underline{Bottom:} Version 2 of the SL-Board.}
\label{fig:tb-desy}
\end{figure}

The new system has been tested with four short SiW ECAL layers with 256 cells each during a beam test in 2019 at DESY. The total setup comprised nine layers. The left part of Figure~\ref{fig:tb-desy} visualises the high level of integration. After a few hours of commissioning on-site the readout system has worked seamlessly for one week until the end of the beam test without a single failure. The graphical interface allowed for quick adjustments of the detector parameters during the running and the simple but yet performant and intuitive online monitoring allowed for a quick feedback of the data quality during the data taking.

During Autumn 2019 a Version 2 of the SL-Board has been designed and produced and is now available, see right part of Figure~\ref{fig:tb-desy}. The main features of this new version are outlined in the following:

\begin{itemize}
\item The length of the short flat cable leaving the SL-Board has been increased from 40 to 48\,mm, which facilitates its connection to the CORE Kapton in the tight environment of a beam test setup;
\item The main input plugs have been rearranged across the board to facilitate the connection of the board to e.g. high voltage and low voltage power supplies;
\item The connector for the FTDI USB module is changed and moved. Now it takes less space and is easier to handle; 
\item A switch permits the encoding the slot number;
\item The SL Board features now a DAC for ADC calibration of the front end ASICs. It also permits delivering calibration pulses with a programmable amplitude. 
\item A flash EEPROM for permanent information storage of e.g. the serial number of connected ASUs;
\item The FPGA produces pulses for autonomous functional calibration the two gains available in the ASIC;
\item The HV is available on the SL-Board to ASU connectors (both sides);
\end{itemize}

Currently a beam test using the digital readout system for up to 15 layers is prepared. This beam test will happen at DESY during 2020. Furthermore, the number of ASUs connected to the readout system and in particular to individual SL-Boards will be increased progressively. The handling of a large(r) number of boards will benefit from the development of the JTAG Interface board. This board has been designed in order to program the SL-Boards firmware through the CORE Module via the Core Kapton.

Since CALICE prototypes share the similar front-end electronic ASICs~\cite{Dulucq:2009nrv}, the readout system can be relatively easily exported to other CALICE prototypes, which is beneficial to the progress of the individual projects and facilitates common beam tests in the coming years. The readout system can also be integrated with beamline instruments such as telescopes. 

\acknowledgments

The authors would like to thank the organisers of the CHEF2019 and LCWS2019 conferences for their hospitality and for giving us the opportunity to present this work.

This work has received funding from the European Union's Horizon 2020 Research and Innovation programme within the AIDA2020 Project under Grant Agreement no. 654168; from the P2IO LabEx (ANR-10-LABX-0038), excellence project HIGHTEC, in the framework Investissements d\'\,Avenir
(ANR-11-IDEX-0003-01) managed by the French National Research Agency (ANR); from the People Programme (Marie
Curie Actions) of the European Union{\textquotesingle}s Seventh Framework Programme (FP7/2007-2013)
under REA grant agreement, PCOFUND-GA-2013-609102, through the PRESTIGE programme coordinated by Campus France.

The measurements leading to these results have been performed at the Test Beam Facility at DESY Hamburg (Germany), a member of the Helmholtz Association (HGF).




\bibliographystyle{JHEP}
\begin{tiny}
\bibliography{mybibfile}
\end{tiny}







\end{document}